\RequirePackage{fix-cm}

\documentclass[smallextended]{svjour3}
\smartqed
\usepackage{comment}
\usepackage{float}
\usepackage{sidecap}
\usepackage{wrapfig}
\usepackage{graphicx}
\usepackage[compatibility=false]{caption}
\usepackage{subcaption}
\usepackage{amssymb,amsmath}
\usepackage{color}
\usepackage{url}
\usepackage{tkz-graph}
\usepackage{algorithm}
\usepackage[noend]{algpseudocode}
\usepackage[justification=centering]{caption}
\usepackage{epstopdf}

\makeatletter
\definecolor{orange}{rgb}{1,0.5,0}
\definecolor{brown}{RGB}{165,42,42}
\makeatother
\newcommand{\hide}[1]{}
\begin{document}

\title{On The Network You Keep: Analyzing Persons of Interest using Cliqster \thanks{A preliminary version of this paper appeared in Proceedings of the 2014 IEEE/ACM International Conference on Advances in Social Networks Analysis and Mining \cite{6921571}.} }

\author{Saber Shokat Fadaee \and
Mehrdad Farajtabar\and
Ravi Sundaram \and
Javed A. Aslam \and
Nikos Passas}

\institute{
S. Shokat Fadaee \at
College of Computer and Information Science \\ Northeastern University \\
\email{saber@ccs.neu.edu} \and M. Farajtabar \at
 College of Computing \\ Georgia Institute of Technology \\ \email{mehrdad@gatech.edu}  \and
R. Sundaram \at College of Computer and Information Science \\ Northeastern University \\ \email{koods@ccs.neu.edu} \and J. A. Aslam \at College of Computer and Information Science \\ Northeastern University \\  \email{jaa@ccs.neu.edu} \and N. Passas \at
School of Criminology and Criminal Justice \\ Northeastern University \\ \email{n.passas@neu.edu}
}

\date{Received: date / Accepted: date}

\maketitle

\begin{abstract}

Our goal is to determine the structural differences between different categories of networks and to use these differences to predict the network category. Existing work on this topic has looked at social networks such as Facebook, Twitter, co-author networks etc.  We, instead, focus on a novel data set that we have assembled from a variety of sources, including law-enforcement agencies, financial institutions, commercial database providers and other similar organizations. The data set comprises networks of \textbf{persons of interest} with each network belonging to different categories such as suspected terrorists, convicted individuals etc. We demonstrate that such ``anti-social'' networks are qualitatively different from the usual social networks and that new techniques are required to identify and learn features of such networks for the purposes of prediction and classification.

We propose Cliqster, a new generative Bernoulli process-based model for unweighted networks. The generating probabilities are the result of a decomposition which reflects a network's community structure. Using a maximum likelihood solution for the network inference leads to a least-squares problem. By solving this problem, we are able to present an efficient algorithm for transforming the network to a new space which is both concise and discriminative. This new space preserves the identity of the network as much as possible. Our algorithm is interpretable and intuitive. Finally, by comparing our research against the baseline method (SVD) and against a state-of-the-art Graphlet algorithm, we show the strength of our algorithm in discriminating between different categories of networks.

\hide{and we describe an interpretable, and efficient algorithm for representing networks within this model.  Our representation preserves the important
underlying characteristics of the network
and is both concise and discriminative.  We demonstrate the discriminative power of our method by comparing to a traditional SVD method as well as a state-of-the-art Graphlet algorithm.  Our results are general in that they
can be applied to ``person of interest'' networks as well as traditional social media networks.}


\end{abstract}


\section{Introduction}

\subsection{Motivation}

The past decade has seen a dramatic growth in the  popularity and importance of social
networks. Technological advancements have made it possible to follow the digital trail
of the interactions and connections among individuals. Much attention has been paid to
the question of how the interaction among individuals contributes to the structure and
evolution of social networks. In this paper we address the related question of identifying the category of a network by looking at its structure. To be more specific, the central problem we tackle is: given a network or a sample of nodes (and associated induced edges) from a network infer the category of the network utilizing only the network structure. For example given different socializing graphs of people with different careers, we are interested in identifying career of a group of people in a given network using only the structural characteristics of their socializing graph. In a mathematical form, let's assume we are given the graphs $G_{1},G_{2}, \cdots, G_{n}$ and another graph $G_{m}$. We would like to find out which graph has the most similar structure to $G_{m}$, and whether $G_{m}$ can be used to reconstruct any of those graphs.

Rather than studying individuals through popular social  networks (such  as Twitter, Facebook, etc.), the presented research is based on  a new data-set  which has been collected through law-enforcement agencies, financial institutions, commercial databases and  other public resources. Our data-set is a collection of networks of \textbf{persons of interest}.
 This approach of building networks from public resources has been successful because it is often easier to infer the
 connections among individuals from widely available resources than through the private
 activities of specific individuals.

\subsection{Dataset and Problem Statement}

Our dataset has been gathered from a variety of public and commercial
sources including the United Nations \cite{un}, World-Check \cite{wc}, Interpol \cite{ip},
Factiva \cite{djf}, OFAC\cite{of}, Factcheck \cite{fc}, RCMP \cite{rcmp}, and various
police websites, as well as other public organizations.  The final dataset was comprised of 700,000 persons of interest with 3,000,000 connections among them \cite{data}.

Except for a few ``mixed'' networks (a network is a connected component) almost all the networks  belong to one of the above 5 categories, i.e. all the nodes in the network belong to one category.
\hide{The central problem we tackle is: given a network or a sample of nodes (and associated induced edges) from a network infer the category of the network utilizing only the network structure.
With the filtered dataset in hand, we proceeded to address the specific question of
an algorithm's ability to infer a particular social category.} Based on our experiments and analyses,  these networks do not demonstrate the common properties of regular social networks such as the famed small world phenomenon \cite{Kleinberg}. As shown in table \ref{table:1} the number of connected components in each category is  large and thus these networks are not small-world.

We extracted some graph structure features from each individual, such as degree and page rank,  then split the data set into a training($80\%$) and a test($20\%$) data set, and ran a supervised learning method (Multinomial logistic regression) on the training data set. After that we compared the actual values of the test set with the prediction results of the regression and came up with $46.89\%$ accuracy for the page rank and $40.61\%$ accuracy for the graph degree. This justifies the quest for new techniques to identify features in the underlying structure of the networks that will enable accurate classification of their categories.

\subsection{Our contributions}

After performing experiments with decomposition methods (and their variants) from  existing literature, we finally discovered a novel technique we call Cliqster -- based on decomposing the
network into a linear combination of its maximal cliques, similar to Graphlet decomposition \cite{azari12} of a network. We compare Cliqster against the traditional SVD (Singular Value Decomposition) as well as state-of-the-art  Graphlet methods.
The most important yardstick of comparison is the discriminating power of the methods.
We find that Cliqster is superior to Graphlet and significantly superior to SVD
in its discriminating power, i.e., in its ability to distinguish between
different categories of persons of interest.
Efficiency is another important criterion and comprises both the speed of the
inference algorithm as well as the size of the resulting representation.
Both the algorithm speed as well as the model size are closely tied
to the dimension of the bases used in the representation. Again, here the
dimension of the Cliqster-bases was smaller than the Graphlet-bases in a
majority of the categories and substantially smaller than SVD in all
the categories.
A third criterion is the interpretability of the model. By using cliques, Cliqster naturally captures interactions between  groups or cells  of individuals and is thus useful for detecting subversive sets of individuals with the potential to act in concert.

In summary, we provide a new generative statistical model for networks with an efficient
inference algorithm. Cliqster is computationally  efficient, and intuitive, and gives interpretable results.
We have also created a new and comprehensive data-set gathered from public and commercial records that has independent value.
Our findings validate the promise of statistics-based technologies for categorizing
and drawing inferences about sub-networks of people entirely through the structure of their network.

The remaining part of the paper is organized as follows. In \S \ref{sec:rel}, we briefly
introduce related work. \S \ref{sec:snm} presents the core of our argument, describing our
network modeling and the inference procedure. In \S \ref{sec:res}, experimental results are
presented demonstrating the effectiveness of our algorithm  on finding an appropriate
and discriminating representation of a social network's structure. At the end of this
section, we present a comprehensive discussion of observations regarding the dataset.
\S \ref{sec:con} draws further conclusions based on this dataset and 
an introductory note on possible directions for future work.

\section{Related Work}
\label{sec:rel}

Significant attention has been given to to the approach of studying criminal
activity through an analysis of social networks \cite{reiss80}, \cite{gla96}, and \cite{Pat08}. \cite{reiss80} discovered that two-thirds of criminals commit crimes alongside another person. \cite{gla96} demonstrated
that charting  social interactions can facilitate an understanding of criminal activity.
\cite{Pat08} investigated the importance of weak ties to interpret criminal activity.

Statistical network models have also been widely studied in order to demonstrate
interactions among people in different contexts. Such network models have been used to
analyze  social relationships, communication networks, publishing activity, terrorist networks,
and protein interaction patterns, as well as many other huge data-sets. \cite{erdHos1959random}
considered random graphs with fixed number of vertices and studied the properties of this model
as the number of edges increases. \cite{gilbert1959random} studied a related version in
which every edge had a fixed probability $p$  for appearing in a network. Exchangeable random
graphs \cite{airoldi2006bayesian} and exponential random graphs \cite{Robins2007173} are other important models. In \cite{bilgic:vast06} they created a toolbox to resolve duplicate nodes in a social network.

The problem of finding roles of a person in a network has been widely studied. In \cite{DBLP:journals/corr/Barta14} they have a link-based approach to this problem. In \cite{distinguishone} they studied how to identify a group of vertices that can mutually verify each other.  The relationship between social roles and diffusion process in a social network is studied in \cite{yang2014rain}. In \cite{DBLP:journals/corr/abs-1305-7006} they combine the problem of capturing uncertainty over existence of edges, uncertainty over attribute values of nodes and identity uncertainty. In \cite{Henderson:2012:RSR:2339530.2339723} they use an unsupervised method to solve the problem of discovering roles of a node in a network.  In \cite{Zhao:2013:ISR:2487575.2487597} they studied how the network characteristic reflect the social situation of users in an online society. In \cite{10.1109/TKDE.2014.2349913} they study the role discovery problem with an assumption that nodes with similar structural patterns belong to the same role. The difference between the works of \cite{Henderson:2012:RSR:2339530.2339723}, \cite{Zhao:2013:ISR:2487575.2487597}, \cite{10.1109/TKDE.2014.2349913} and similar works like \cite{6729527}, \cite{DBLP:journals/corr/abs-1101-3291}, \cite{node-classification-factor-graph} with our work is that they are interested in the roles of a node in a specific network, while we are interested in studying the structural differences among different networks. In this work, we assume all the nodes in a network has the same role/job. Despite the various applications of finding the roles of different sub networks in a graph, this problem has only received a limited amount of attention. In this paper we are studying the role discovery problem for a network.

Recently researchers have become interested in stochastic block-modeling and latent graph
models \cite{nowicki2001estimation,airoldi08,karrer2011stochastic}. These methods attempt
to analyze the latent community structure behind interactions.
Instead of modeling the community structure of the network directly, we propose a simple
stochastic process based on a Bernoulli trial for generating networks. We implicitly consider
the community structure in the network generating model through a decomposition and
projection to the space of baseline communities (cliques in our model).
For a comprehensive review of statistical network models we refer interested readers to
\cite{survey09}.

Formerly, Singular Value Decomposition was used for the decomposition of a network
\cite{chung96,hoff09,kim12}. However, since SVD basis elements are not interpretable in terms of community structure, it can not capture the notion of social information we are interested in quantifying. Authors in \cite{azari12} introduced the Graphlet decomposition of a weighted network; by abandoning the orthogonality constraint they were able to gain interpretability. The resulting method works with weighted graphs; however, alternate techniques, such as power graphs (which involve powering the adjacency matrix of a graph to obtain a weighted graph), need  to be used in order to apply this method to unweighted graphs such as (most) social networks.

\section{Statistical Network Modeling}
\label{sec:snm}
\subsection{Model}
Let's assume we have $n$ nodes in the network (For example $n = 10$ in Figure 1).
Consider $Y$ as a $n \times n$ matrix representing the connectivity in the network.
$Y(r,s) = 1$ if node $r$ is connected to node $s$, and $0$ otherwise.

\begin{figure}[H]
\centering
\includegraphics[width=80mm]{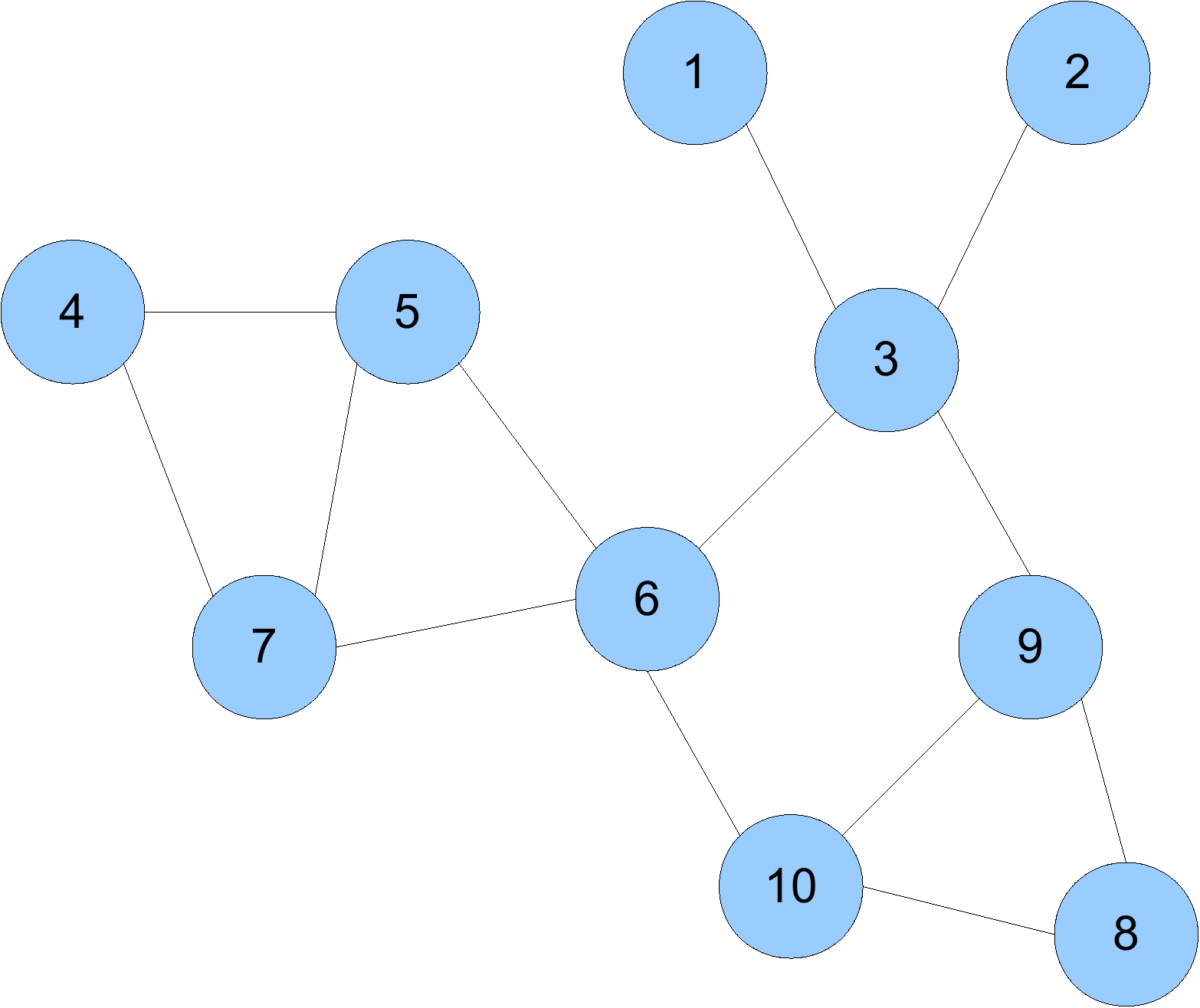}
\caption{Network of ten people}
\label{fig:b1}
\end{figure}

%

%

In Cliqster, the generative model for the network is:
\begin{equation}
Y = \text{Bernoulli}(Z)
\end{equation}
which means $Y(r,s) = Y(s,r) =1$ with probability $Z(r,s)$, and $Y(r,s)=Y(s,r) = 0$ with probability $1-Z(r,s)$ for all $r > s$. Since the graph is undirected the matrix $Z$ is lower triangular.

Inspired by PCA and SVD, in Cliqster we choose to represent $Z$ in a new space \cite{chung96}, \cite{kim12}. Community structure is a key factor to understand and analyze a network, and because of this we are motivated to choose bases in a way that reflects the community structure \cite{hoff09}. Consequently, we decided to factorize $Z$ as
\begin{equation}
Z = \sum_{k=1}^{K} \mu_k B_k
\end{equation}
where $K$ is the number of maximal cliques (bases), and $B_k$ is $k^{th}$ lower triangular basis matrix that represents the $k^{th}$ maximal clique, and $\mu_k$ is its contribution to the network.
In section \ref{basis_sec} we elaborate on this basis selection process. From this point forward, we consider these bases as cliques of a network. We also represent a network in this new space. Each network is parameterized by the coefficients and the bases which construct the $Z$, the network's generating matrix.

\subsection{Inference}
When given a network $Y$ of people and their connections, our goal is to infer the parameters generating this network.
We must first assume the bases are selected as baseline cliques. The likelihood of the network parameters (coefficients) given the observation is:
\begin{equation*}
\mathcal{L}(\mu_{1:K}) =
\prod_{r>s : Y(r,s)=1} Z(r,s)
\prod_{r>s : Y(r,s)=0} (1-Z(r,s))
\end{equation*}
We estimate these parameters by maximizing their likelihood under the constraint $0 \le Z(r,s) \le 1$ for all $r>s$.

One can easily see the likelihood is maximized when $Z(r,s)=1$ if $Y(r,s)=1$ and $Z(r,s)=0$ if $Y(r,s)=0$. Therefore
\begin{equation}
Y = \sum_{k=1}^K \mu_k B_k
\end{equation}
should be used for the lower triangle of $Y$.

Unfolding the above equation results in,
\begin{equation*}
\begin{split}
& Y(2,1) = \mu_1 B_1(2,1) + \ldots + \mu_K B_K(2,1) \\
& Y(3,1) = \mu_1 B_1(3,1) + \ldots + \mu_K B_K(3,1) \\
& Y(3,2) = \mu_1 B_1(3,2) + \ldots + \mu_K B_K(3,2) \\
& \vdots \\
& Y(n,n-1) = \mu_1 B_1(n,n-1) + \ldots + \mu_K B_K(n,n-1)
\end{split}
\end{equation*}
We define two vectors,
\begin{equation}
\begin{split}
& \boldsymbol{\mu} = (\mu_1, \ldots, \mu_K)^{\top} \\
& \boldsymbol{b^{rs}} = (B_1(r,s) , \ldots, B_K(r,s))^{\top}
\end{split}
\end{equation}
So the new objective function can  be written as,
\begin{equation}
J = \sum_{r>s} (\boldsymbol{\mu}^{\top} \boldsymbol{b^{rs}} - Y(r,s))^2
\end{equation}

$J$  is convex with respect to $\mathbf{\mu}$ under the following constraints $0 \le \boldsymbol{\mu}^{\top} \boldsymbol{b^{rs}}  \le 1$. This is essentially a constrained least squares problem, which can be solved through existing efficient algorithms \cite{lawson1974solving}, \cite{boyd2004convex}. Through this formula, the representation parameters $\mu_{1:K}$ are thus computed easily and we are done with the inference procedure.

We turn our attention to the new representation and try to find an algorithm which can produce a more interpretable result. The exact generating parameters are no longer needed in our application. Therefore, by relaxing the constraints we will be able to present it with a simple and very efficient algorithm. In addition, the solution to this unconstrained problem provides us with an intuitive understanding of what is happening behind this inference procedure. To determine the optimal parameters, we must take the derivative with respect to $\mu$:
\begin{equation}
\frac{\partial J}{\partial \mathbf{\mu}} =
2 \sum_{r>s} \boldsymbol{b^{rs}} ({\boldsymbol{b^{rs}}}^{\top} \boldsymbol{\mu} - Y(r,s))
\end{equation}
By equating the above derivative to zero and doing a simple mathematical procedure, we are presented with the solution
\begin{equation}
\boldsymbol{\mu} = A^{-1} \boldsymbol{d}
\end{equation}
where
\begin{equation}
\begin{split}
& A =  \sum_{r>s} \boldsymbol{b^{rs}} \boldsymbol{b^{rs}}^{\top} \\
& \boldsymbol{d} = \sum_{r>s} Y(r,s) \boldsymbol{b^{rs}}
\end{split}
\end{equation}

$A$ is a $K \times K$ matrix and $\boldsymbol{d}$ is a $K \times 1$ vector. Thus, while we still have a very small least squares problem, it has been significantly reduced from the original equation in which there were $O(n^2)$ constraints. Despite this fact, we obtain very good results, and we will soon explain why this happens.

Our novel decomposition method finds $\mu$ which is used to represent a network, and which could stand-in for a network in network analysis applications. This representation is used in the next section in order to discriminate between different types of networks.

The results from the decomposition of the network presented in  figure 1  is demonstrated in table \ref{table:mu1}.

\begin{table}[h]
\small
\caption{$\mu$ within each cluster}
\label{table:mu1}
\begin{center}
\begin{tabular}{lcc}
{\bf Cluster members} &{\bf $\mu$}   \\
\hline \\
$\{ 8, 9, 10 \}$ &1.00 \\
$\{ 5, 6, 7 \}$  &0.75\\
$\{ 4, 5, 7 \}$ &0.75\\
$\{ 1, 2, 3 \}$ &1.00 \\
$\{ 6, 10 \}$ &1.00\\
$\{ 3, 9 \}$ &1.00\\
$\{ 3, 6 \}$ &1.00\\
\end{tabular}
\end{center}
\end{table}

\subsection{Interpretation}
In general, it is not an easy task to interpret the Eigenvectors of an SVD. In our model, however, all the values of $A$ and $\boldsymbol{d}$ give you an intuition about the network. For further insight into this process, consider a matrix $A$. Every entry of this matrix is equal to the number of edges shared by the two corresponding cliques.
This matrix encodes the power relationships between baseline clusters, as a part of network reconstruction. The intersection between two bases shows how much one basis can overpower another basis as they are reconstructing a network. In contrast, $\boldsymbol{d}$ presents the commonalities between a given network and its baseline communities. Through this equation, a community's contribution to a network is encoded.

With the interpretation of this data in mind, the equation $A \boldsymbol{\mu} = \boldsymbol{d}$ is now more meaningful for understanding the significance of our new representation of a network. Consider multiplying the first row of the matrix by the vector $\boldsymbol{\mu}$, which should be equal to $\boldsymbol{d}_1$. In order to solve this equation, we have chosen our coefficients in such a way that when the intersection of cluster 1 and other clusters are multiplied by their corresponding coefficients and added together, the result is a clearer understanding of the first cluster's contribution to the network construction.

\subsection{Basis Selection}
\label{basis_sec}
Users in \emph{persons of interest} network usually form associations in particular ways, thus, community structure is a good distinguishing factor for different networks.  There are different structures that form a community. One of the interesting structures that forms a community is the maximal cliques of that community. We use them as the basis of our method.
There are so many ways to compute the maximal cliques of a network. We use the Bron-Kerbosch algorithm \cite{bron73}  for identifying our network's communities. As mentioned in \cite{azari12}, this is one of the most efficient algorithms for identifying  all of the maximal cliques in an undirected network. After applying the Bron-Kerbosch algorithm to figure 1, we identify the communities that are represented in table \ref{table:1}. The Bron-Kerbosch algorithm is described in the algorithm \ref{alg:bron}.

\begin{algorithm}
\caption{Bron-Kerbosch algorithm}
\label{alg:bron}
\begin{algorithmic}[1]
\State $C = \emptyset$
\Comment{We keep the maximal clique in C}
\State $I = V(G)$
\Comment{The set of vertices that can be added to C}
\State $X =  \emptyset$
\Comment{The set of vertices that are connected to C but are excluded from it}

\Procedure{Enumerate}{$C, I, X$}
\If{$I == \emptyset$ and $X == \emptyset$}
\State $C$ is maximal clique
\Else
\For{each vertex v in I}
\State $Enumerate(C \cup \{v\}, I \bigcap N(v), X \bigcap N(v))$
\State           $I \gets I \ \{v\}$
 \State $X \gets X \cup \{v\}$

\EndFor
\EndIf
\EndProcedure
\end{algorithmic}
\end{algorithm}

The Bron-Kerbosch algorithm has many different versions. We use the version introduced in \cite{eppstein2011listing}.

One of the most successful aspects of this algorithm is that it provides a multi-resolution perspective of the network. This algorithm identifies communities through a variety of scales, which, we will see, allows us to locate the most natural and representative set of coefficients and bases.

\subsection{Complexity}
\label{Complexity}

The aforementioned inference equation requires $A$ and $\boldsymbol{d}$ to be computed, which can be done in $O(m + n)$ time where $m$ is the number of edges and $n$ is the number of nodes in the network.
The least-square solution requires $O(K^3)$ operations.
A graph's degeneracy measures its sparsity and is the smallest value $f$ such
that every nonempty induced subgraph of that graph contains a vertex of degree at most $f$ \cite{lick70}.
In \cite{eppstein2011listing} they proposed a variation of
the Bron-Kerbosch algorithm, which runs in $O(f n 3^{f/3})$ where $f$ is a network's degeneracy number.
This is close to the best possible running time since the largest possible number of maximal cliques in an n-vertex graph with degeneracy $f$ is $(n - f) 3^{f/3}$ \cite{eppstein2011listing}. 

A power law graph is a graph in which the number of vertices with degree $d$ is proportional to $x^{\alpha}$ where $1 \leq \alpha \leq 3$. When $1 < \alpha \leq 2$ we have $f = O(n^{1/2\alpha})$, and when $2 < \alpha < 3$ we have $f = O(n^{(3-\alpha)/4})$ \cite{buchanan13}. Combining with the running time, $O(f n 3^{f/3})$ of the Bron-Kerbosch variant \cite{eppstein2011listing}, we find that the running time for finding all maximal cliques in a power law graph to be $2^{O(\sqrt{n})}$.

However, the maximum number of cliques in graphs based on real world  networks is typically $O(\log n)$ \cite{azari12}.

\section{Results}
\label{sec:res}
In this section we investigate the properties of the new features we have learned about the network  in question. Firstly, we introduce the new dataset we have built.
Our experiments attempt to prove two claims:
\begin{enumerate}
\item the new representation is concise, and
\item it can discriminate between different network types
\end{enumerate}
We will now compare our results with SVD decomposition and graphlet decomposition algorithms \cite{azari12}.

\subsection{Dataset}

We have gathered a  dataset by gathering and fusing information from a variety of public and commercial sources. Our final dataset was comprised of around 750,000 persons of interest with 3,000,000 connections among them. We then filtered this dataset to slightly less than 550,000 individuals who fell
into one of the following 5 categories:
\begin{enumerate}
\item \emph{Suspicious Individuals}: Persons who have appeared on sanctioned lists, been arrested or detained, but not been convicted of a crime.
\item \emph{Convicted Individuals}: Persons who have been indicted, tried and convicted in a court of law.
\item \emph{Lawyers/Legal Professionals}: Persons currently employed in a legal profession.
\item \emph{Politically Exposed Persons}: Elected officials, heads  of parties, or persons who
have held or currently hold political positions now or in the past.
\item \emph{Suspected Terrorists}: Persons suspected of aiding, abetting or committing terrorist activities.
\end{enumerate}

This dataset is publicly available at \cite{data}.

\begin{table}[H]
\small
\caption{Table of Categories and corresponding sizes plus number of connected components and density of each category}
\label{table:1}
\begin{center}
\begin{tabular}{llll}
{\bf Category}  &{\bf Members} &{\bf Components} &{\bf Density} \\
\hline \\
Suspicious Individuals         &316,990 &77,811 &  0.0000180 \\
Convicted Individuals             &165,411 &35,517& 0.0000427 \\
Lawyers/Legal Professionals              &3,723 &1,492 & 0.0006220\\
Politically Exposed Persons &13,776 &4,947 & 0.0001533\\
Suspected Terrorists &31,817 &5,016 & 0.0002068\\
\end{tabular}
\end{center}
\end{table}

The color scheme we use for our figures are as follow: \textcolor{red}{Red} for \textcolor{red}{\textbf{Suspicious Individuals (SI)}}, \textcolor{blue}{blue} for \textcolor{blue}{\textbf{Convicted Individuals (CI)}}, \textcolor{brown}{brown} for \textcolor{brown}{\textbf{Lawyer/Legal Professionals (LL)}}, \textcolor{orange}{orange} for \textcolor{orange}{\textbf{Politically Exposed Persons (PEPS)}}, and \textbf{black} for \textbf{Suspected Terrorists (ST)}
.  

\subsection{Basic properties}

\begin{figure}
\centering
\includegraphics[width=100mm,height = 85mm]{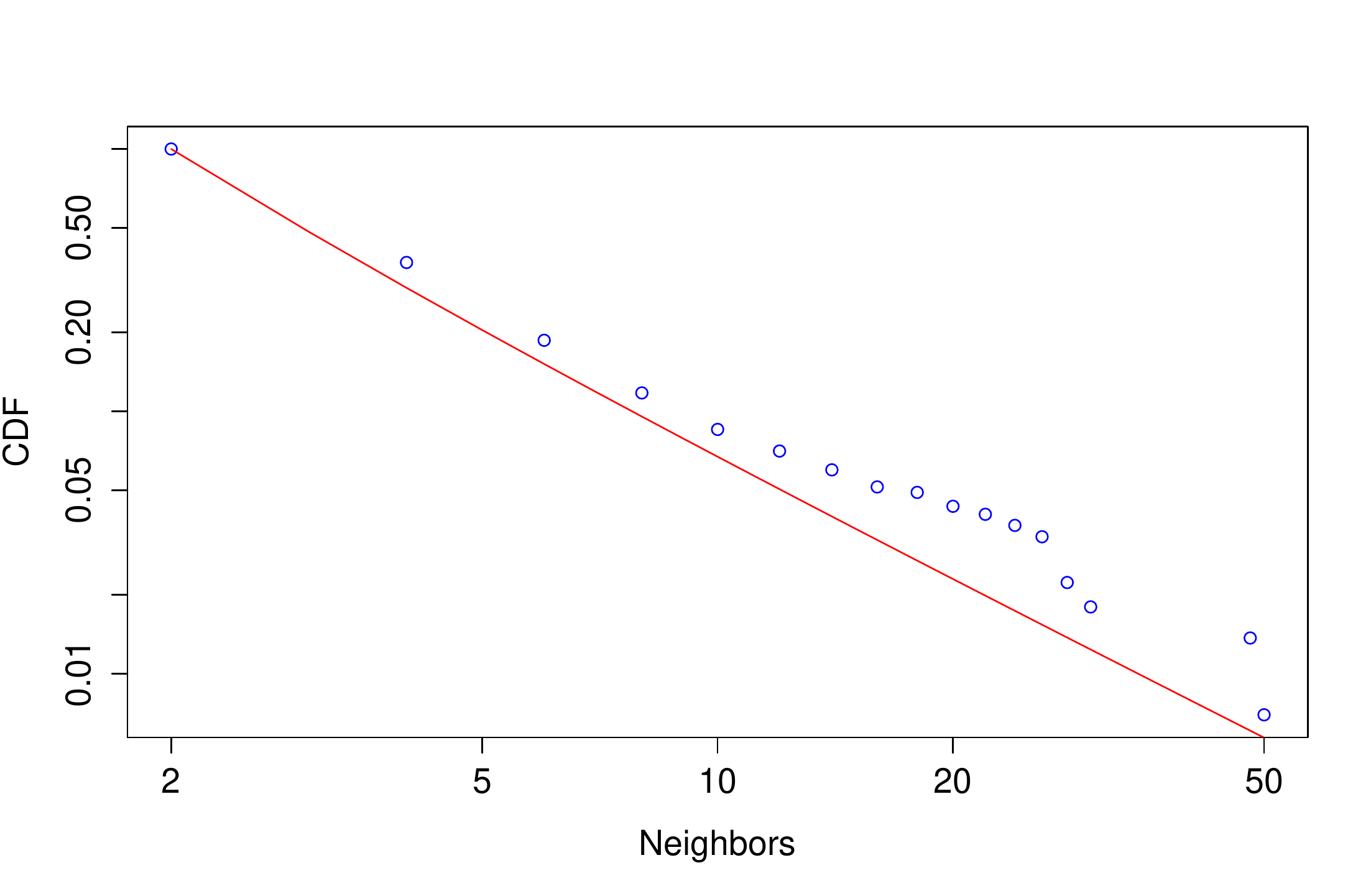}
\caption{The cumulative distribution functions and their maximum likelihood power law fit of the Legal subnetwork}
\label{fig:loglog}
\end{figure}

We want to know whether our dataset has the common properties of social networks or not, i.e. having a power law distribution. The first thing to check is the degree distribution of each subnetwork, and if they can be fitted to a power-law distribution. We have a scale-free network If the degree distributions in our subnetwork follow power-law distribution. We used the poweRlaw \cite{powerlaw} and igraph \cite{igraph} packages to calculate the maximum likelihood power law fit of the Legal subnetwork, and the results are shown in figure \ref{fig:loglog}.   It looks like a scale-free network, but we need to check this with more accurate measures. In a power-law distribution $P(X = x)$ is proportion to $c x^{\alpha}$. The $\alpha$ of each subnetwork can be seen in the table \ref{table:alpha}. Each of our subnetwork can be fitted into a power-law distribution, so all of them are scale-free networks. However, these networks are not small-world networks. The number of connected components in each network, indicates if you start at a certain node in each network it is impossible to reach to most of the other nodes in that network.

\begin{table}[H]
\small
\caption{Table of alpha, the exponent of the fitted power-law distribution in each category}
\begin{center}
\begin{tabular}{ll}
{\bf Category}  &{\bf $\alpha$}  \\
\hline \\
Suspicious Individuals         &1.838563  \\
Convicted Individuals             &1.733839 \\
Lawyers/Legal Professionals              &2.977307 \\
Politically Exposed Persons &3.107326 \\
Suspected Terrorists &1.770715 \\
\end{tabular}
\end{center}
\label{table:alpha}
\end{table}

\subsection{Sampling method}
For each category we choose a random induced sub-graph of a $1,000$  vertices as a sample.
We then analyze this data, and repeat this operation $1,000$ times and represent the data's
average with bold lines in the following graphs. All figures also include a representation of
what happens to this data when the standard deviation of it is taken at a margin of 2 , which we
illustrate through a line of a lighter variation of the same color. We analyzed this data with
three different methods, the Singular Value Decomposition, Graphlet Decomposition, as well as
our own proposed model.

\subsection{Singular Value Decomposition}
\begin{figure}
\centering
\includegraphics[width=90mm]{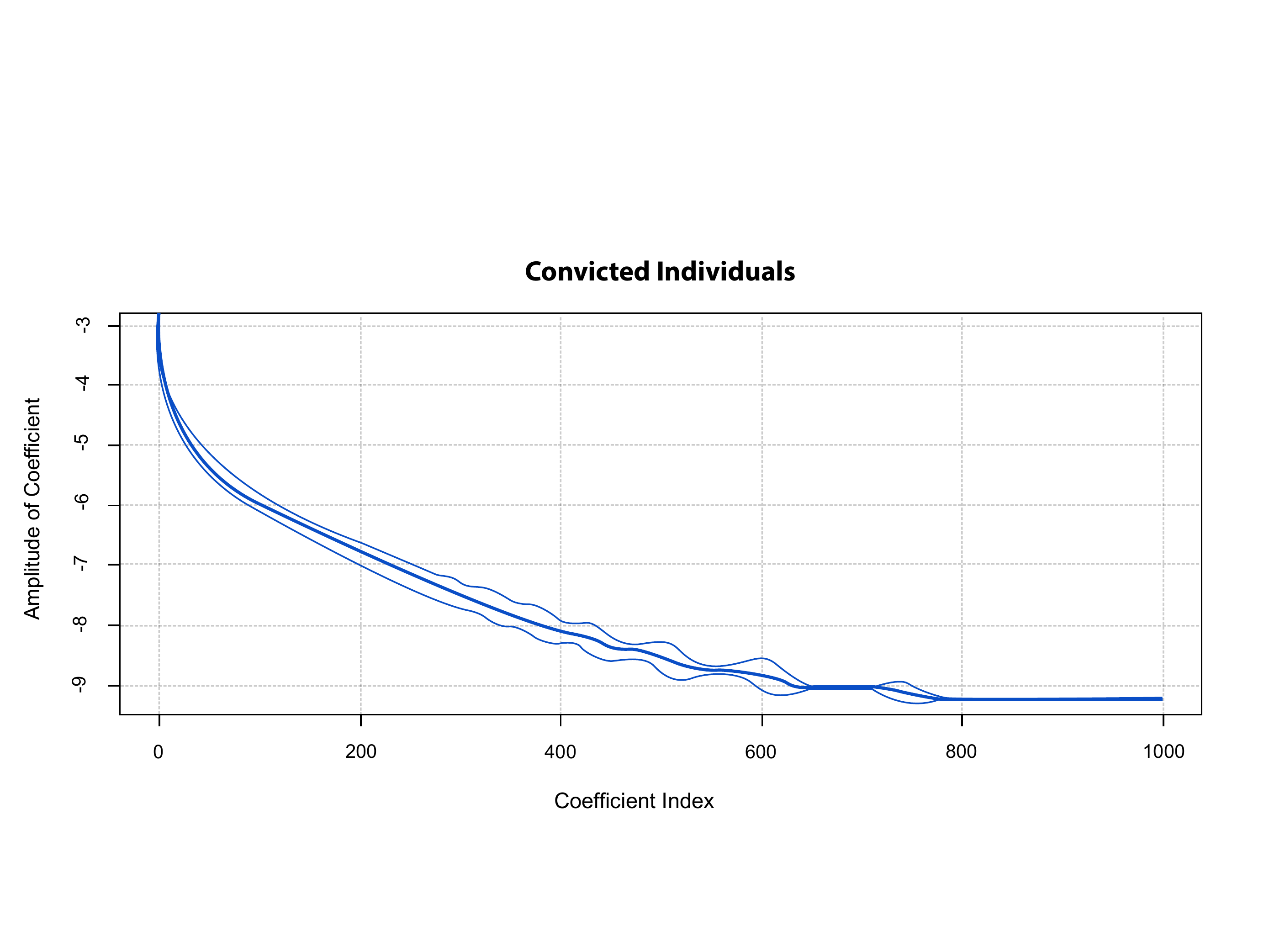}
\caption{Number of bases and amplitude of coefficient for Convicted Individuals using SVDNumber of bases and amplitude of coefficient for Convicted Individuals using SVD}
\label{fig:e1}
\end{figure}

\begin{figure}
\centering
\includegraphics[width=90mm]{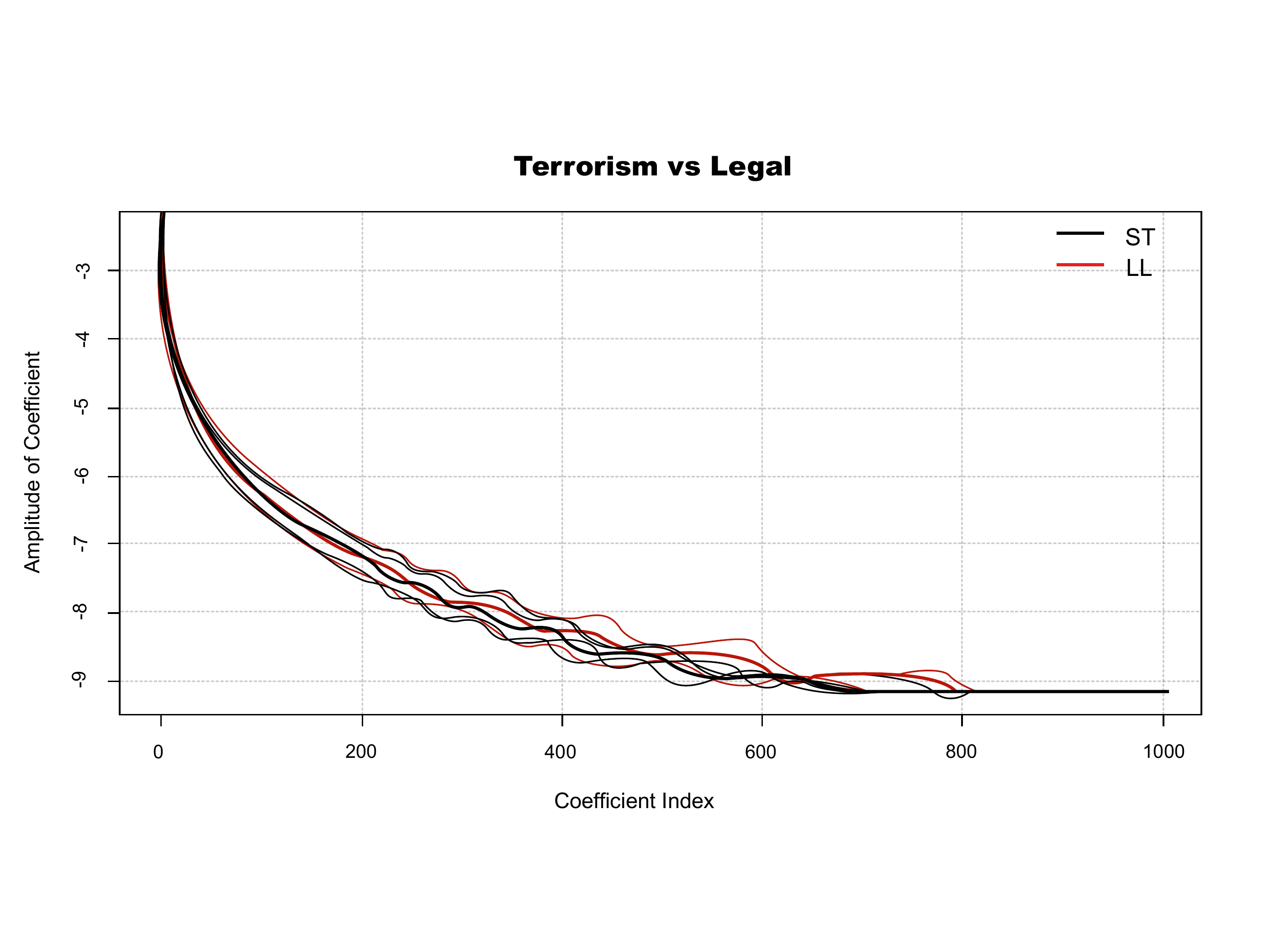}
\caption{Comparison of coefficients between Terrorist sub networks and Legal sub networks using SVD}
\label{fig:e2}
\end{figure}

\begin{figure}
\centering
\includegraphics[width=90mm]{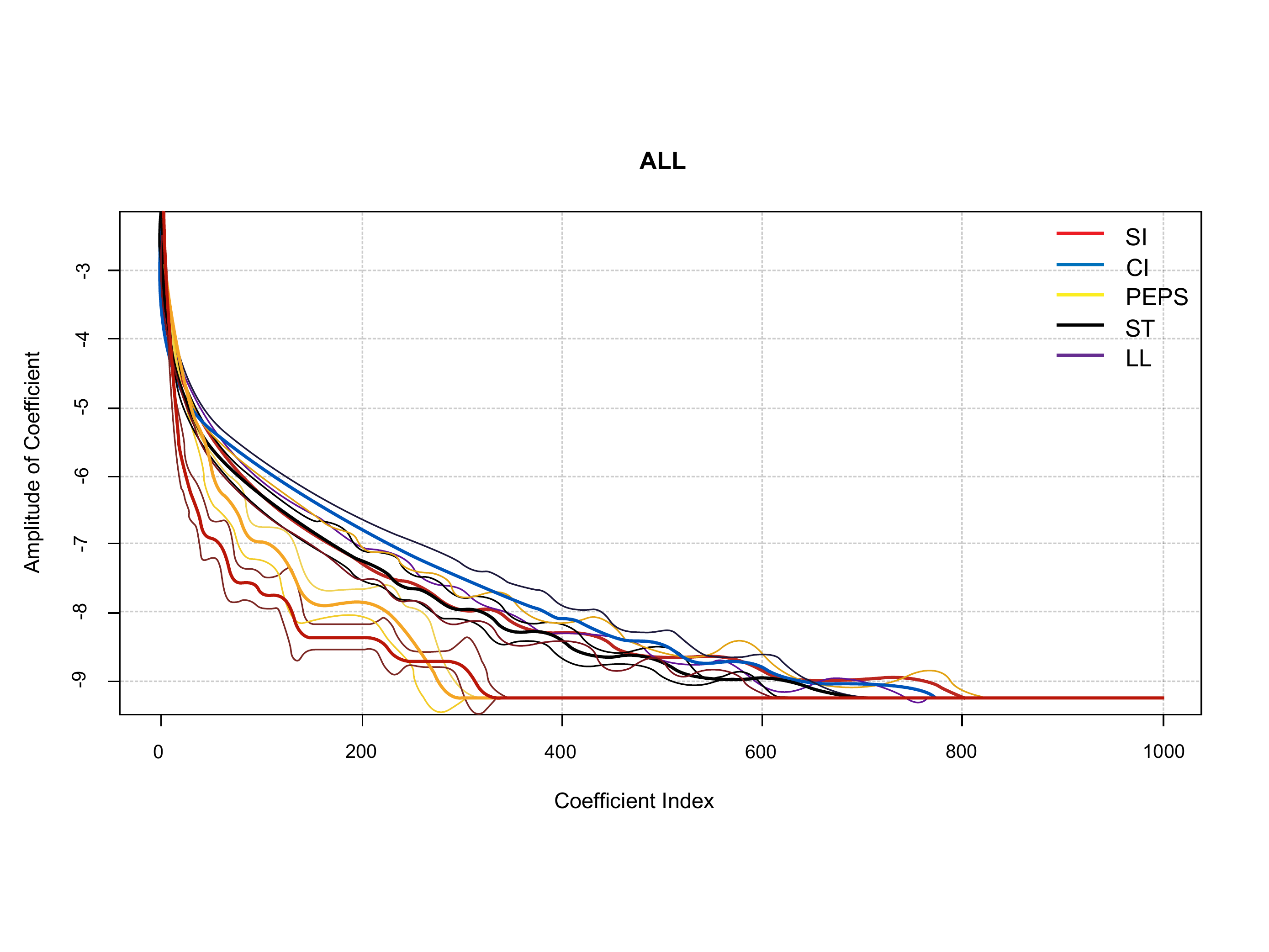}
\caption{The ability of SVD method to distinguish between different categories of networks}
\label{fig:e3}
\end{figure}

We first analyzed our data using the Singular Value Decomposition method \cite{chung96}. Figure \ref{fig:e1}  shows the effective number of non-zero coefficients for this algorithm. Figure \ref{fig:e2} demonstrates the ability of this algorithm to discriminate between two different categories. Finally, the ability of the algorithm to distinguish between the 5 categories is illustrated in figure \ref{fig:e3}. The average number of bases we observed in the samples of a $1,000$ vertices is around $800$ as can be seen in figures \ref{fig:e1}, \ref{fig:e2} and \ref{fig:e3}.

\subsection{Graphlet Decomposition}

\begin{figure}
\centering
\includegraphics[width=90mm, height=100mm]{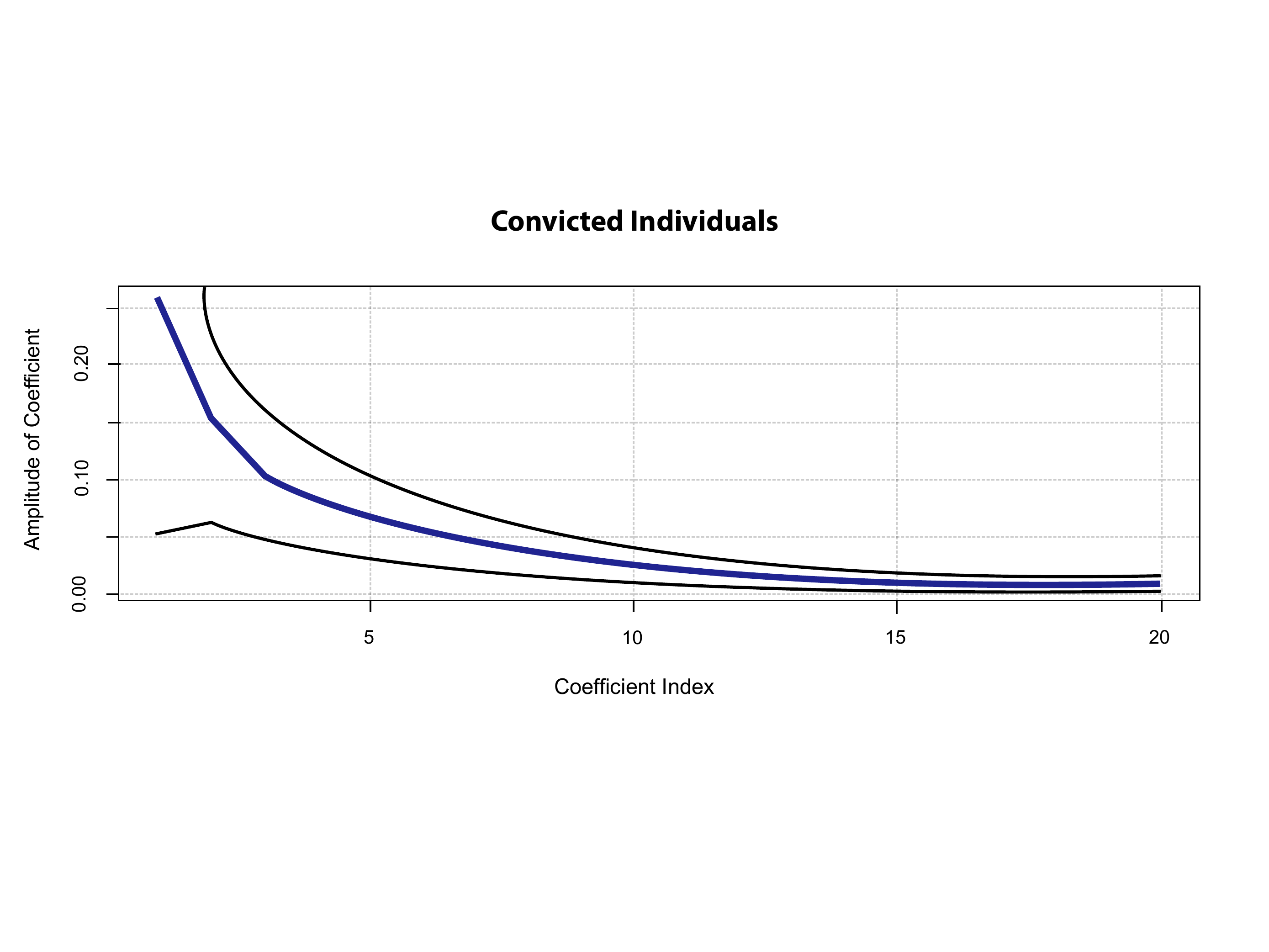}
\caption{Number of bases and amplitude of coefficient for Convicted Individuals using Graphlet Decomposition Algorithm}
\label{fig:g1}
\end{figure}

\begin{figure}
\centering
\includegraphics[width=90mm, height=100mm]{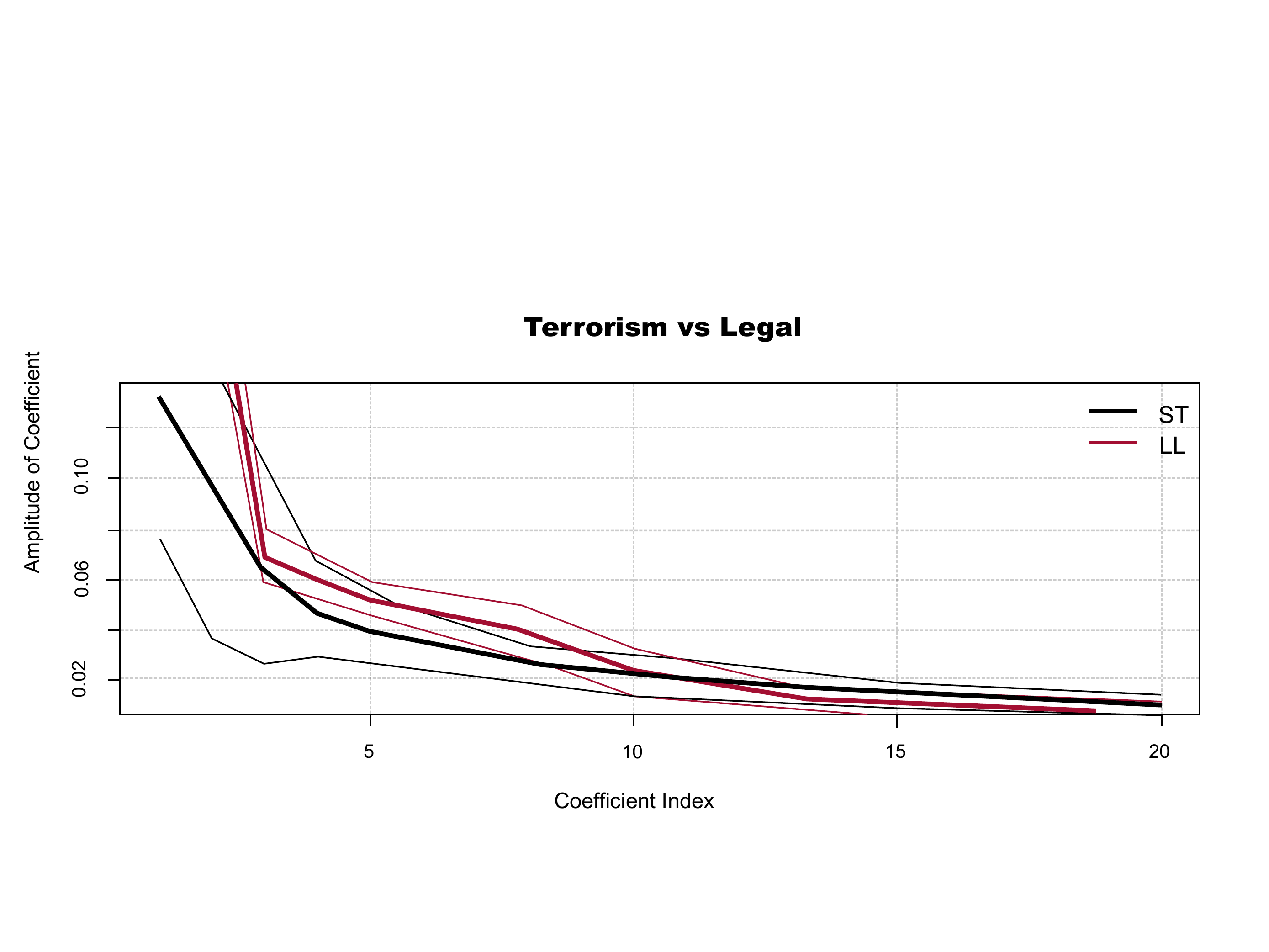}
\caption{Comparison of coefficients between Terrorist sub networks and Legal sub networks using Graphlet Decomposition Algorithm}
\label{fig:g2}
\end{figure}

\begin{figure}
\centering
\includegraphics[width=90mm, height=100mm]{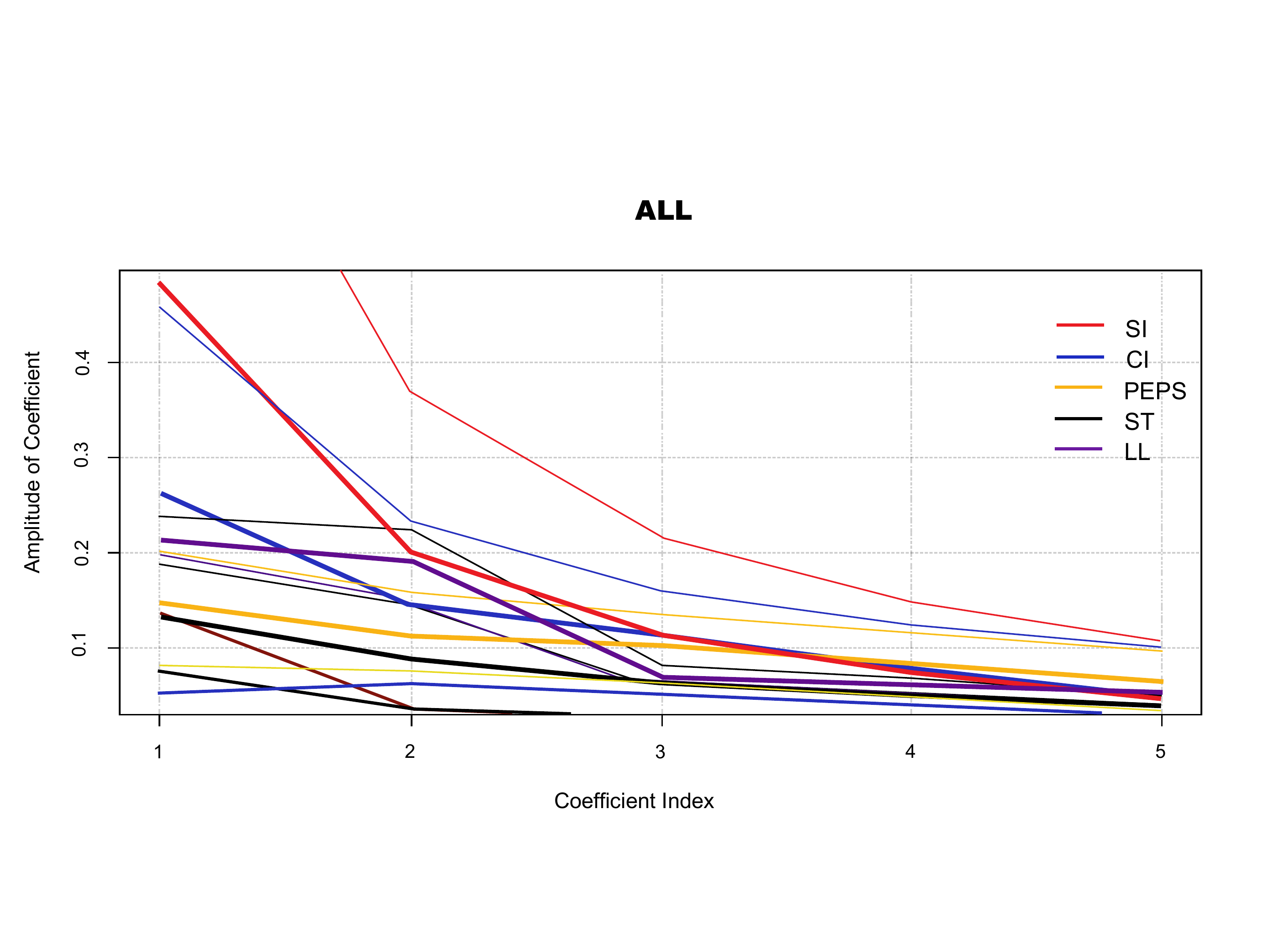}
\caption{The ability of Graphlet method to distinguish between different categories of networks}
\label{fig:g3}
\end{figure}

We next performed the same tests using Graphlet Decomposition. Figure \ref{fig:g1} demonstrates the effective number of non-zero coefficients for this algorithm. Figure \ref{fig:g2} shows the ability of this algorithm to discriminate between two different types of networks. The algorithm's ability to distinguish between the 5 categories is again illustrated in figure \ref{fig:g3}.
As can be seen in these figures the number of bases elements for Graphlet Decomposition is around $20$.

\subsection{Cliqster}

\begin{figure}
\centering
\includegraphics[width=90mm]{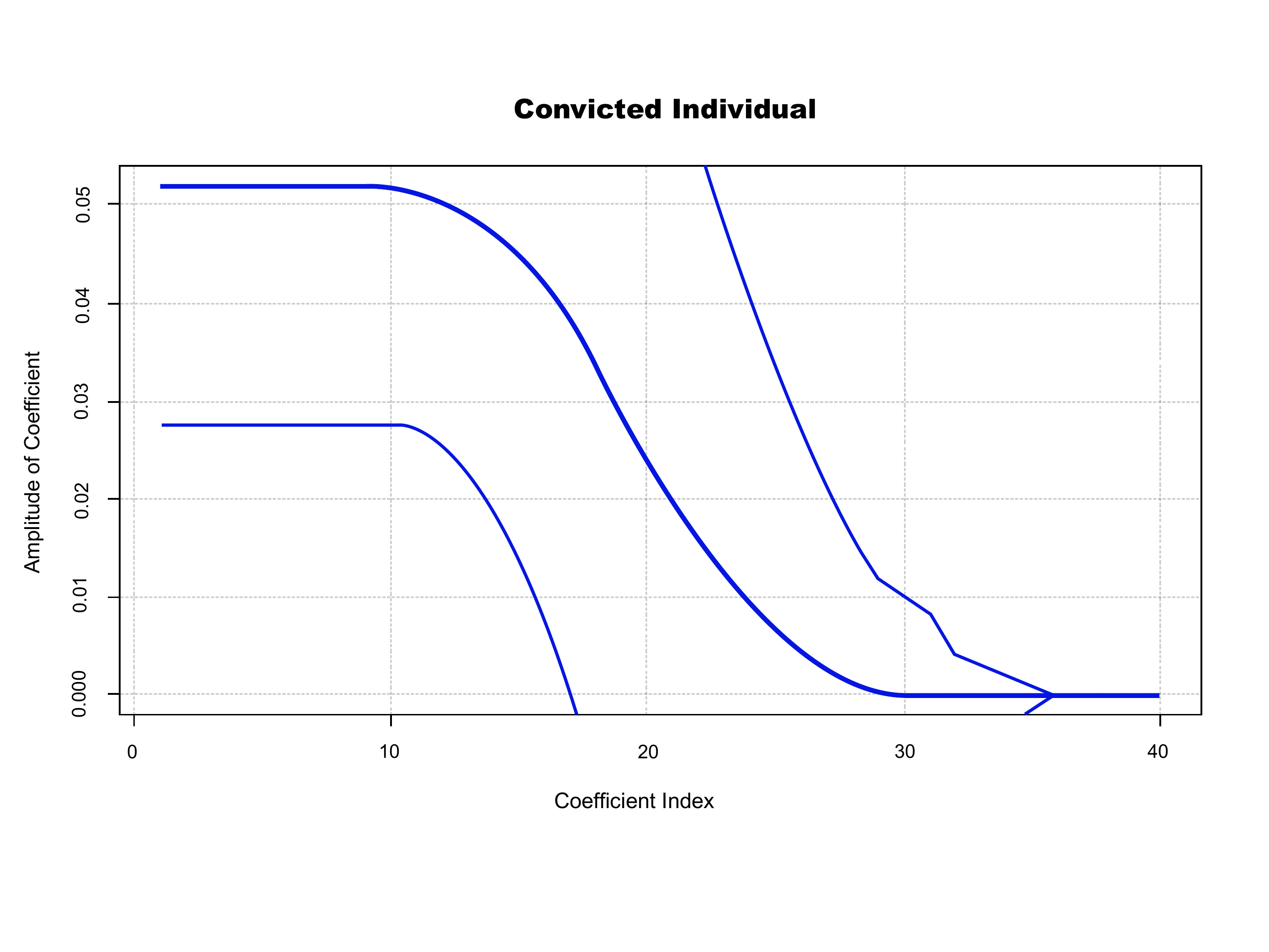}
\caption{Number of bases and amplitude of coefficient for Convicted Individuals using Cliqster}
\label{fig:o1}
\end{figure}

\begin{figure}
\centering
\includegraphics[width=90mm, height=90mm]{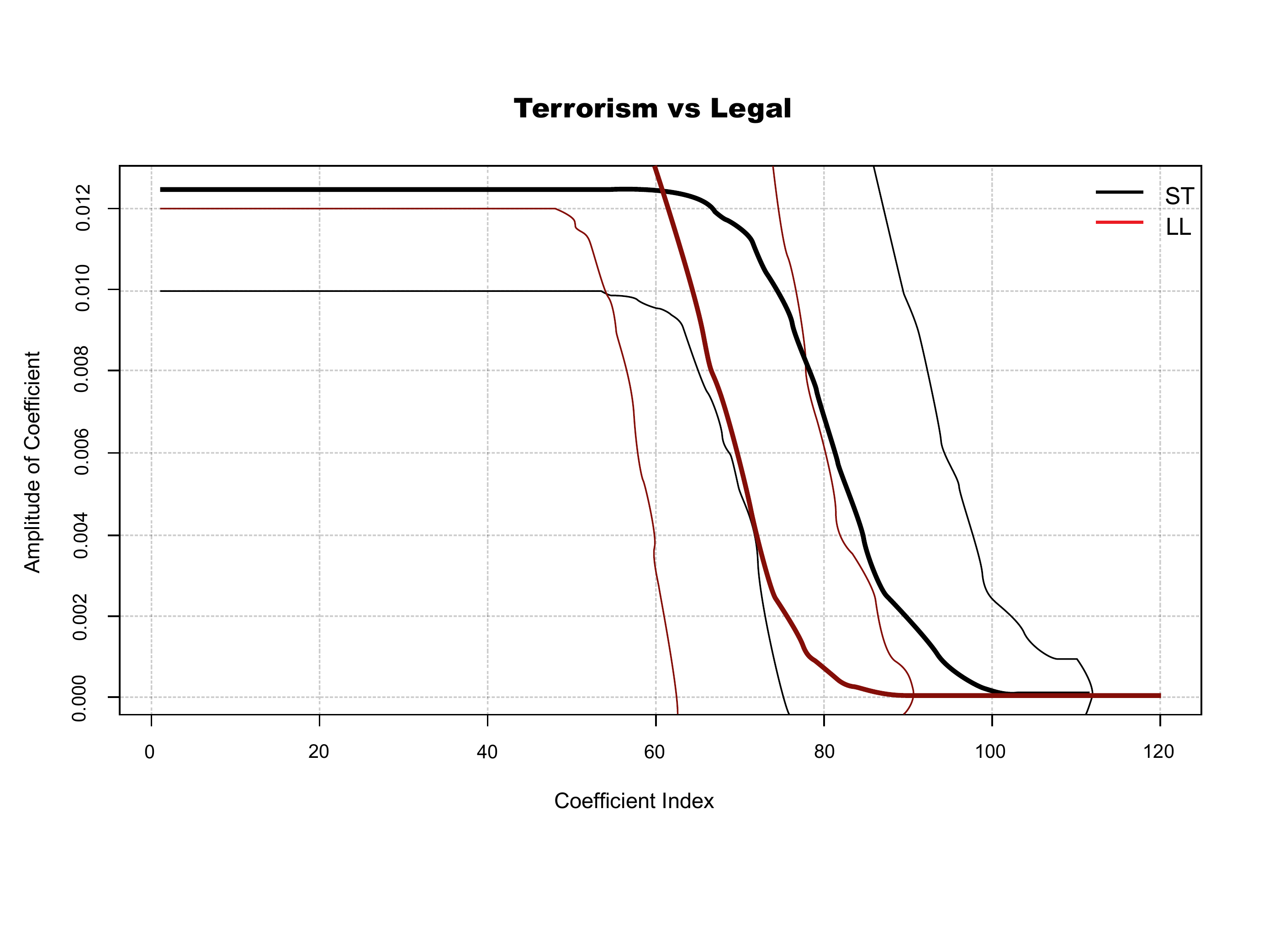}
\caption{Comparison of coefficients between Terrorist sub networks and Legal sub networks using Cliqster}
\label{fig:o2}
\end{figure}

\begin{figure}
\centering
\includegraphics[width=90mm, height=90mm]{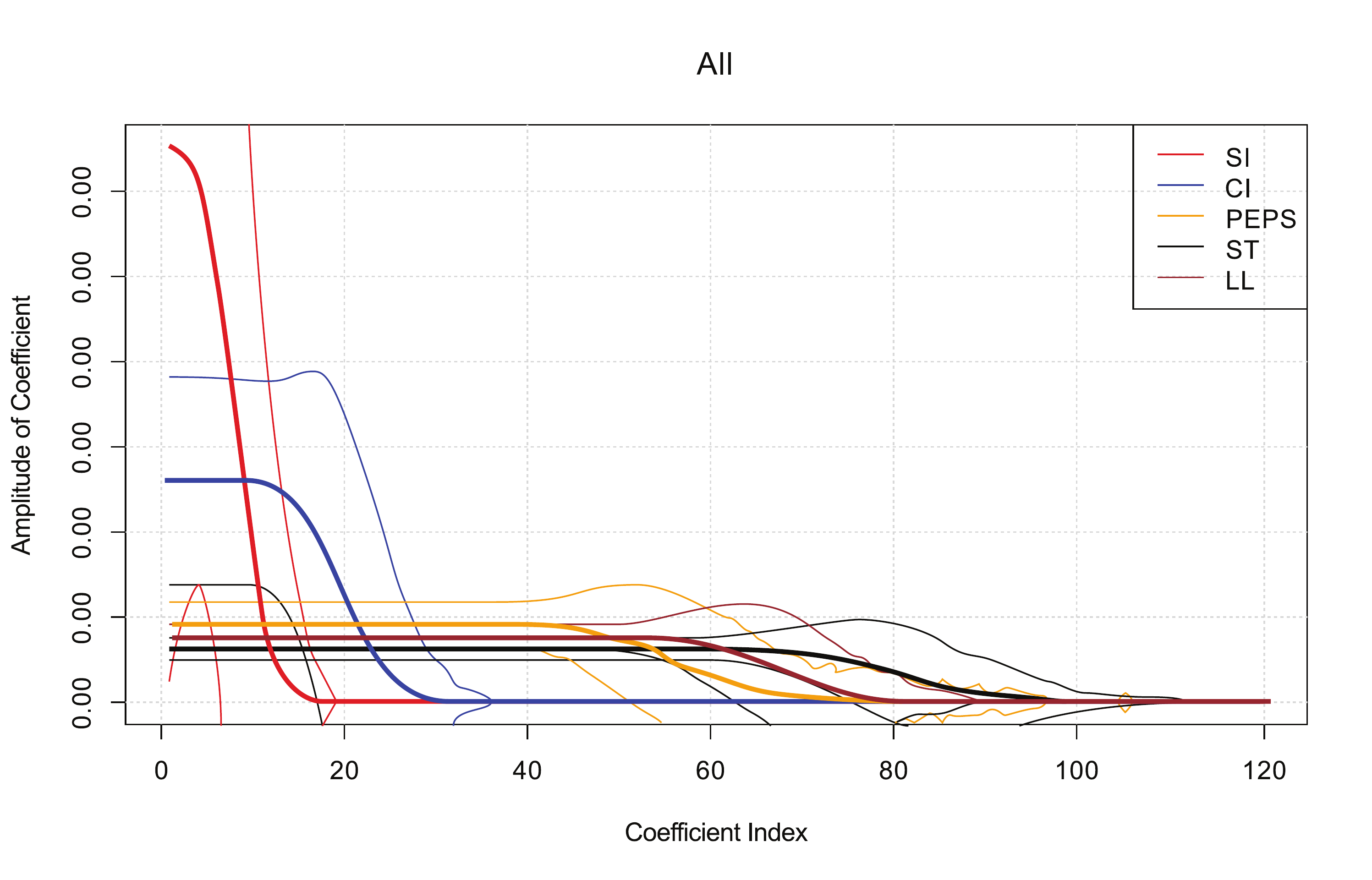}
\caption{The ability of Cliqster to distinguish between different categories of networks}
\label{fig:o3}
\end{figure}

Finally, we performed the same tests using our method. We first determined appropriate bases using the Bron-Kerbosch algorithm. We then computed $A$ and $\boldsymbol{d}$. The new representation for a sample network of one category that resulted from our new method is shown in Figure \ref{fig:o1}. Figure \ref{fig:o2} shows the ability of our algorithm to discriminate between two different types of networks. Our new algorithm's ability to distinguish between two different types of networks is illustrated in Figure \ref{fig:o3}, which also shows that the number of bases elements for Graphlet Decomposition is around $50$.

\subsection{Performance}
\label{perf_sec}
We analyzed the time complexity of Cliqster in the section \ref{Complexity}. Now it's time to check if the empirical results verify our theory. For the \textit{Convicted Individuals} subnetwork we ran both our method and SVD using the igraph package in R. The performance of the Graphlet method is very similar to Cliqster so we do not include that in this experiment.

\begin{figure}
\centering
\includegraphics[width=90mm, height = 80mm]{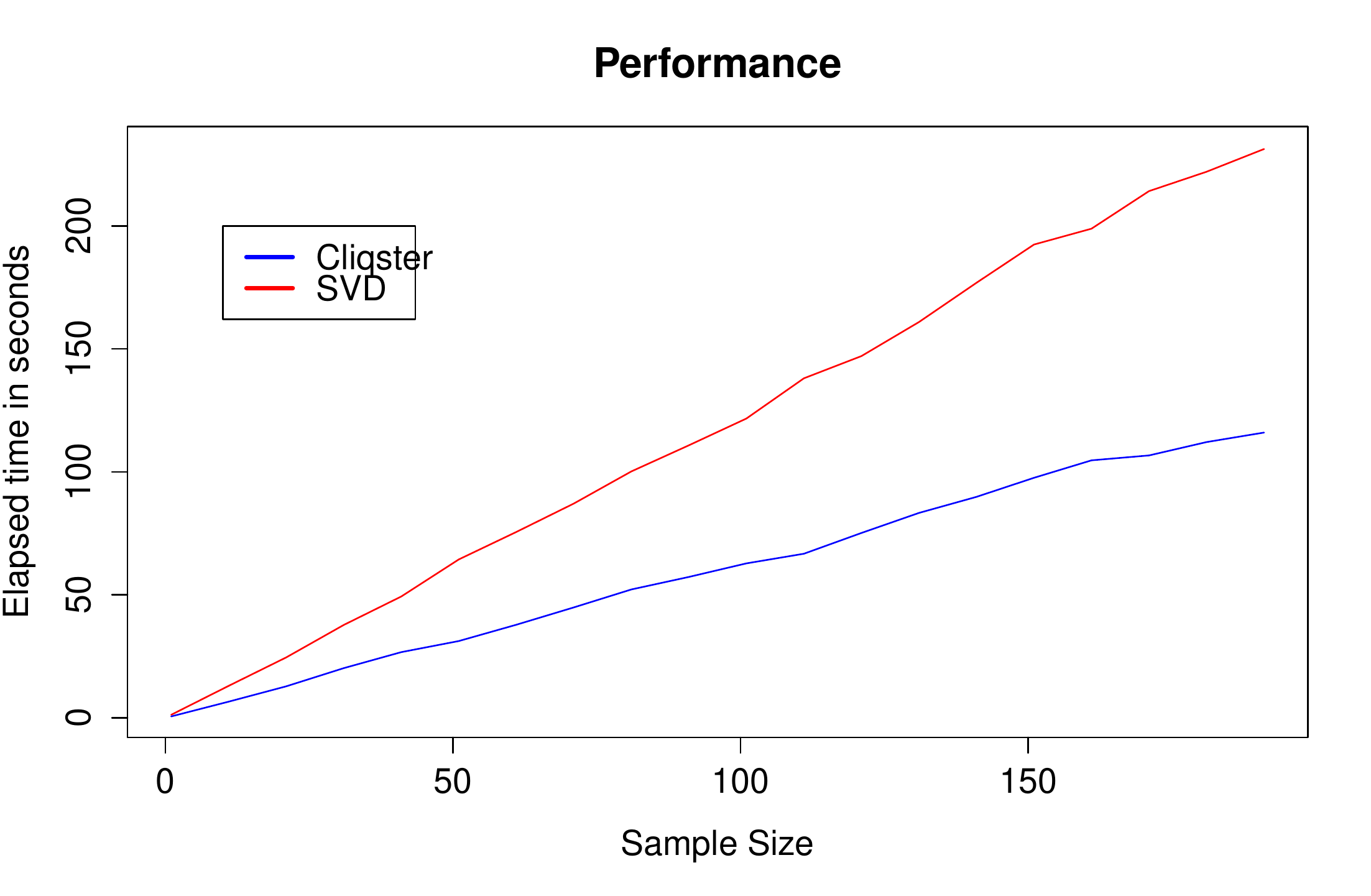}
\caption{Comparison of performance between Cliqster and SVD}
\label{fig:perf}
\end{figure}

We ran our experiment on  ``\textit{Intel(R) Core(TM) i7-2600 CPU @ 3.40GHz (8 CPUs), ~3.4GHz}'' processor with ``\textit{16384MB}'' of memory. As you can see in figure \ref{fig:perf}, as we grow the sample size our method performs twice as fast as the SVD method.

\subsection{Distinguishability}
\label{dist_sec}

In order to compare the ability of each of these methods to distinguish between different types of social networks, we sampled $100$ networks from each category, combining all of these samples before running the K-means clustering algorithm (with $5$  as the number of clusters), and repeated this action $100$ times. We used each network's top $20$ largest coefficients, and are willing to know if coefficients of different sub-networks can be distinguished from each other. We gave the combined coefficients of all different sub-networks to the K-means clustering algorithm as an input, and calculated the mean error of clustering. As you can see in table \ref{table:2}, our method often returns the  bases with the best ability to distinguish between the type of social network presented. The Graphlet Decompostion slightly outperforms our method in two of the following sub-networks, and such difference is  negligible in practice.

\begin{table}[H]
\small
\caption{Mean error of clustering with $20$  coefficients ($\mu_{1:20}$) } \label{table:2}
\label{dist_table}
\begin{center}
\begin{tabular}{lccc}
{\bf Category} &{\bf SVD}  &{\bf Graphlet}  & {\bf Cliqster}\\
\hline \\
SI &0.51461 & \bf{0.00817} & 0.0177\\
CI  &0.71080 &0.11535 & \bf{0.0141}\\
LL  &0.75006 &0.10931 & \bf{0.0153} \\
PEPS &0.66082 &0.12195 & \bf{0.0114} \\
ST &0.65381 & \bf{0.01303} &0.0176 \\
\end{tabular}
\end{center}
\end{table}

\subsection{Classification}

Another method for checking the ability of Cliqster to produce the features that can distinguish  between different networks, is to use $k-$nearest neighbors algorithm (or $k-NN$ for short). $k-NN$ is a non-parametric method that is used for classification in a supervised setting. Let's assume we want to compare the features that are used to distinguish between these two groups: Suspicious Individuals and Convicted Individuals. We train Cliqster with samples of size $1,000$ that are randomly selected from both communities, gather the features and repeat this operation $1,000$ times. After that we run the $k-NN$ with $k=3$ and a test data of size $100$.  In order to avoid ties, we need to pick an odd number for $k$ in case of binary classification. When we set $k=3$ we are looking at the classification problem in a 3 dimensional space. We also make sure there is no intersection between the members of training and test sets to avoid the problem of over-fitting.

\begin{figure}
\centering
\includegraphics[width=90mm, height = 80mm]{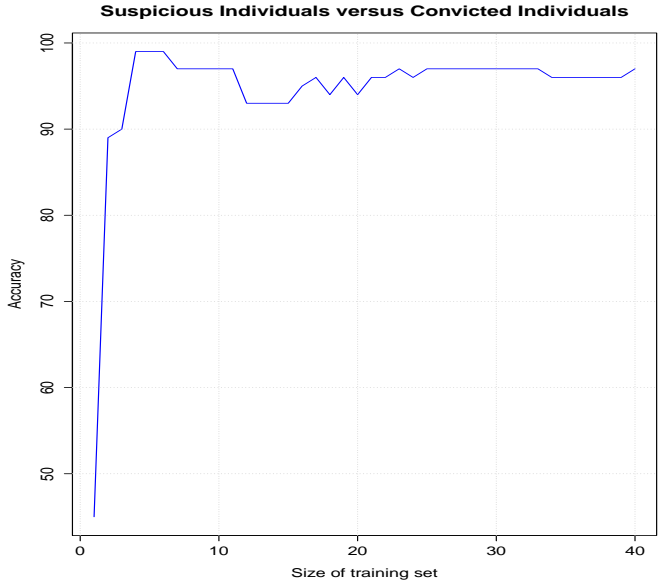}
\caption{The accuracy of community detection based on the training size}
\label{fig:criVSind}
\end{figure}

Figure \ref{fig:criVSind} shows the result of this experiment. With using a training set of size $40$ we can classify these two groups with an accuracy of $97\%$. It basically means that when we have a training set of size 40, K-NN can learn how to distinguish between these two groups with an accuracy of $97\%$.

\begin{figure}
\centering
\includegraphics[width=90mm, height = 80mm]{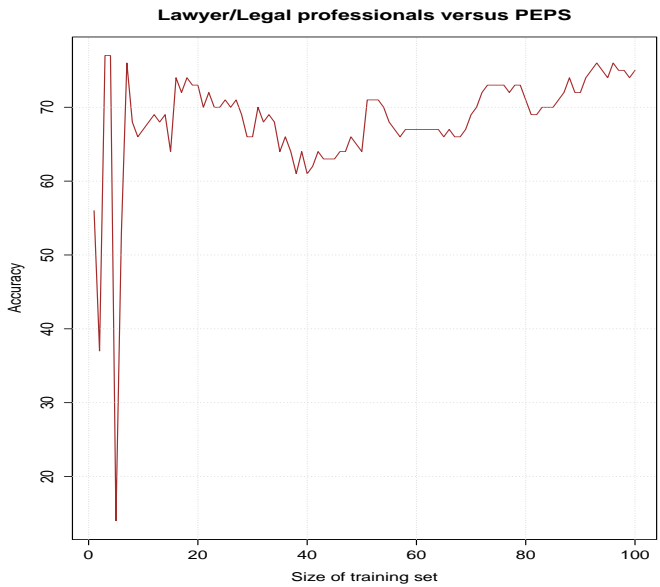}
\caption{The accuracy of community detection based on the training size}
\label{fig:legVSpol}
\end{figure}

Things are a little bit different when it comes to comparing the behavior of Lawyers/Legal professionals network and Politically Exposed Persons network. As you can see in figure \ref{fig:legVSpol} we need a training set of size $100$ to reach to an accuracy of $74\%$. This difference suggest a contrast between the characteristics of these networks.  According to Cliqster,  the network structure of Lawyers/Legal professionals and the network structure of Politically Exposed Persons have more in common than the network structure of Suspicious Individuals and the network structure of Convicted Individuals. 

\begin{figure}
\centering
\includegraphics[width=90mm, height = 80mm]{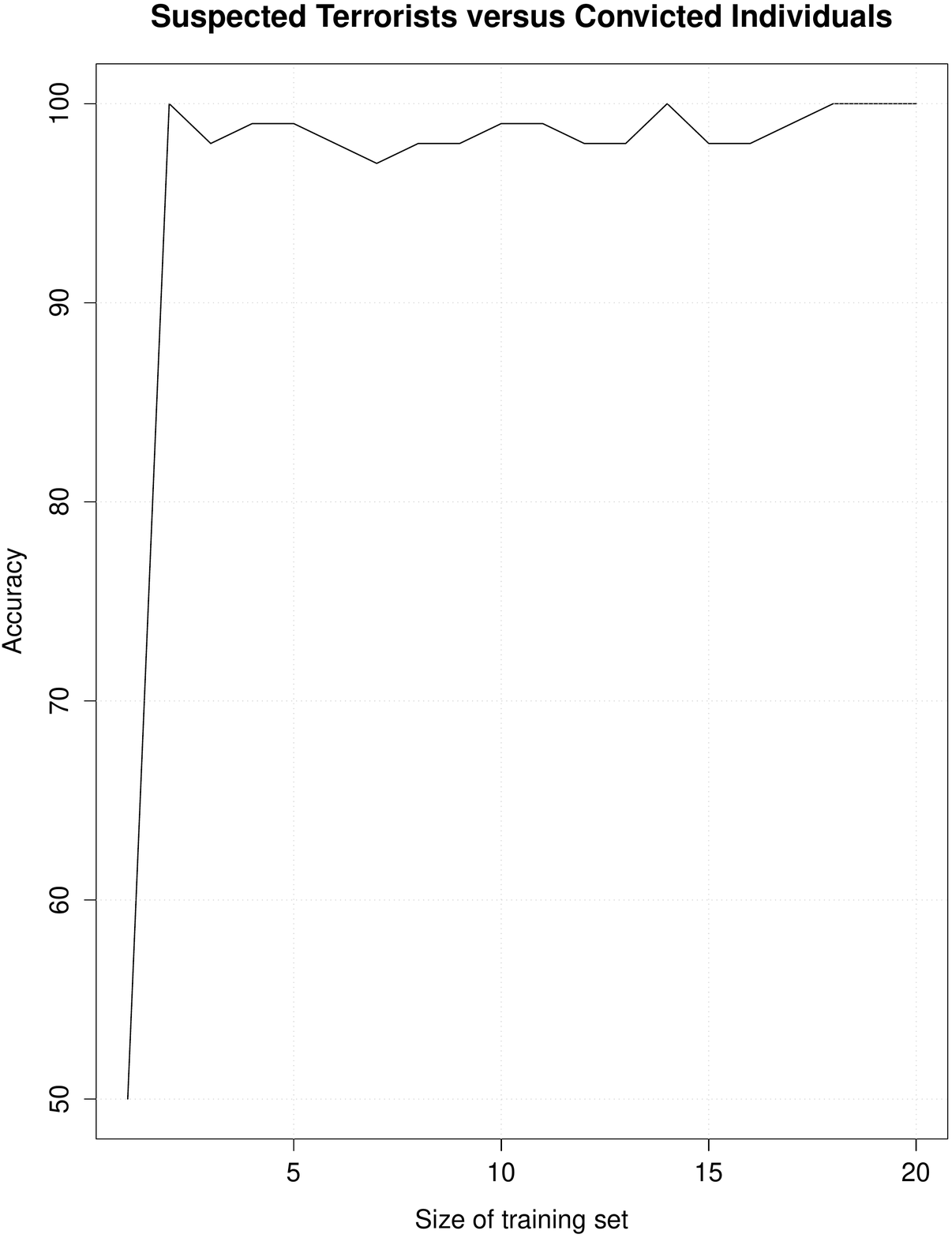}
\caption{The accuracy of community detection based on the training size}
\label{fig:terVScri}
\end{figure}

\begin{figure}
\centering
\includegraphics[width=90mm, height = 80mm]{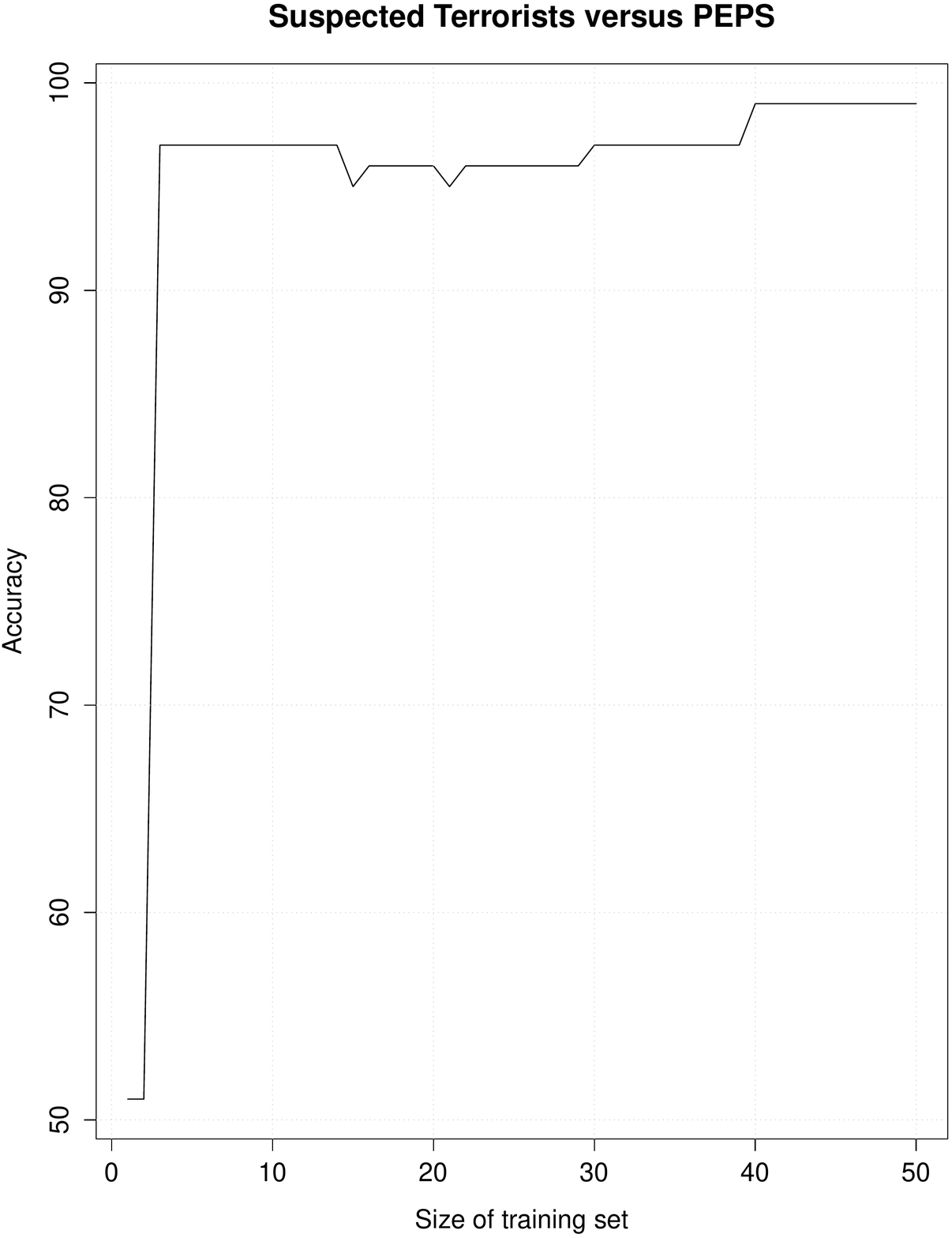}
\caption{The accuracy of community detection based on the training size}
\label{fig:terVSpol}
\end{figure}

If we analyze the network structure of Suspected Terrorists and compare it with network structure of Convicted Individuals, we will see that after using a training set of size around $20$ we reach to the $100\%$ accuracy. $k-NN$ can classify these two groups with no error \ref{fig:terVScri}. Now we compare the network structure of Suspected Terrorists  and Politically Exposed Persons networks \ref{fig:terVSpol}. After using a training set of size $50$, we reach to the $99\%$ accuracy.

\subsection{Discussion}
Figures \ref{fig:e1}, \ref{fig:g1}, and \ref{fig:o1} compare the ability of the three methods to compress data. These graphs demonstrate that the SVD method is inefficient for summarizing a network's features. The graph also shows that the Graphlet method produces the smallest feature space. Our representation is also very small, however, and the difference in size produced through these methods is negligible in real world applications of this equation. Earlier we demonstrated that the $20$ largest coefficients in the representation produced through our method is sufficient to outperform the Graphlet algorithm in terms of distinguish ability and clustering.

Figures \ref{fig:e2}, \ref{fig:g2}, and \ref{fig:o2} demonstrate the ability of the algorithms to distinguish between two selected categories. When comparing our method with the SVD and Graphic Decomposition methods, the coefficients seem to be very similar between those produced by our method and the SVD method, however, our method also performs as well as the Graphlet Decomposition method in distinguishing between two types of networks. This demonstrates that community structure is a natural basis for interpreting social networks. By decomposing a network into cliques, our method provides an efficient transformation that is concise and easier to analyze than SVD bases, which are constrained through their requirement to be orthogonal.
Figures \ref{fig:e3}, \ref{fig:g3}, and \ref{fig:o3} verify these claims for all  5 categories.

Table \ref{dist_table} demonstrates the performance of our algorithm to consistently summarize each network according to category. We then clustered all  coefficients using k-means. Through this process, it became clear that the SVD method could not identity the category of the network being analyzed.
Because of this, we can infer that by selecting the community structure (cliques) as bases, our ability to identify a network is considerably improved. Our proposed algorithm was more accurate in clustering than the Graphlet Decomposition algorithm. Thus, the Bernoulli Distribution (as used in seminal work of Erd{\H{o}}s and R{\'e}nyi) is a simpler and more natural process for generating networks. Our proposed method is also easier to interpret and does not run the risk of getting stuck in local minima like the Graphlet method.

Finally, figures \ref{fig:criVSind}, \ref{fig:legVSpol}, \ref{fig:terVSpol} and \ref{fig:terVScri} demonstrate the ability of $k-NN$ to classify features produced by Cliqster in binary classification settings. They also give us some interpretations on similarities and differences between the network structure of different groups.

\section{Conclusion}
\label{sec:con}
After proposing Cliqster, which is a new generative model for decomposing random networks, we applied this method to our new dataset of persons of interest.
Our primary discovery in this research has been that a variant of our decomposition method provides a statistical test capable of accurately discriminating between different categories of social networks.
Our resulting method is both accurate and efficient.
We created a similar discriminant based on the traditional Singular Value Decomposition and Graphlet methods, and found that they are not capable of discriminating between social network categories. Our research also demonstrates community structure or cliques to be a natural choice for bases. This allows for a high degree of compression and at the same time preserves the identity of the network very well. The new representation produced through our method is concise and discriminative.

Comparing the three methods, we found that the dimensions of the Graphlet-bases and our bases were significantly smaller than the SVD-bases, while also accurately identifying the category of the network being analyzed. Therefore, our method is an extremely accurate and efficient means of identifying different network types.

On the non-technical side we would like to see how we can get law-enforcement agencies to adopt our methods. There are a number of directions for further research on the technical front.
We would like to expand the use of our simple intuitive algorithm to weighted networks, such as networks with an edge generating process based on the Gamma distribution.
The problem with the Maximum Likelihood solution for a network is that it is subject to over-fitting or a biased estimation. Adding a regularization term would adjust for this discrepancy. A natural choice for such a term would be a sparse regularization, which is in accordance with real social networks. Extensive possibility for future work exists in the potential of incorporating prior knowledge into Cliqster by using Bayesian inference. Another natural avenue for further investigations is to consider how Cliqster can be adapted to regular social networks.

\section*{Acknowledgment}

The authors would like to thank Hossein Azari Soufiani for his comments on different aspects of this work.

\bibliographystyle{IEEEtran}
\bibliography{biblio}

\end{document}